\documentclass[manuscript]{aastex}
\usepackage{natbib}
\usepackage{color}
\shorttitle{HXR and Field changes During Flares}
\shortauthors{Burtseva et al.}

\begin{document}

\title{Hard X-ray Emission During Flares \\ and Photospheric Field Changes}

\author{O. Burtseva\altaffilmark{1}, J. C. Mart\'{i}nez-Oliveros\altaffilmark{2}, G. J. D. Petrie\altaffilmark{1}, and A. A. Pevtsov\altaffilmark{1}}
\email{burtseva@noao.edu}
\altaffiltext{1}{National Solar Observatory, Tucson, AZ 85719}
\altaffiltext{2}{Space Sciences Laboratory, UC Berkeley, Berkeley, CA 94720}

\begin{abstract}
We study the correlation between abrupt permanent changes of magnetic field during X-class flares observed by the GONG and HMI instruments, and the hard X-ray (HXR) emission observed by RHESSI, to relate the photospheric field changes to the coronal restructuring and investigate the origin of the field changes. We find that spatially the early RHESSI emission corresponds well to locations of the strong field changes. The field changes occur predominantly in the regions of strong magnetic field near the polarity inversion line (PIL). The later RHESSI emission does not correspond to significant field changes as the flare footpoints are moving away from the PIL. Most of the field changes start before or around the start time of the detectable HXR signal, and they end at about the same time or later than the detectable HXR flare emission. Some of the field changes propagate with speed close to that of the HXR footpoint at a later phase of the flare. The propagation of the field changes often takes place after the strongest peak in the HXR signal when the footpoints start moving away from the PIL, i.e. the field changes follow the same trajectory as the HXR footpoint, but at an earlier time. Thus, the field changes and HXR emission are spatio-temporally related but not co-spatial nor simultaneous. We also find that in the strongest X-class flares the amplitudes of the field changes peak a few minutes earlier than the peak of the HXR signal. We briefly discuss this observed time delay in terms of the formation of current sheets during eruptions.
\end{abstract}

\keywords{Sun: flares - Sun: magnetic fields - Sun: photosphere - Sun: X-rays}

\section{Introduction}

Significant, abrupt and permanent photospheric magnetic field changes during major X-class and M-class flares have been reported by several authors \citep[e.g.,] [] {Wang92,Wang94,Kosovichev01,Wang02,Sudol05,Petrie10,Wang12,Petrie13}. No numerical model has yet reproduced these observations and physically related the field changes at the photosphere (where the magnetic field is not force-free) with the flare that occurs in the nearly force-free field environment in the corona. \cite{Fletcher08} suggested a scenario when the Alfv\'{e}n wave pulses transport energy rapidly from the flare site in the corona to the lower atmosphere as a possible explanation of the observed rapid variations to the line-of-sight (LOS) component of the photospheric magnetic field during the flare impulsive phase. \cite{Hudson08} introduced the loop-implosion scenario, \cite{Hudson08} and \cite{Fisher12} provided the method for estimating flare-related Lorentz force back-reaction, and several studies \citep[e.g.,] [] {Wang10,Petrie10,Wang12,Alvarado12,Petrie12,Petrie13} estimated the Lorentz forces from magnetic field observations. \cite{Petrie14} showed that the method of \cite{Fisher12} could be used for coherent changes of strong fields within active regions.

\cite{Johnstone12} analyzed the spatial and temporal relationship between the photospheric field changes and chromospheric ultraviolet (UV) emission. They found that the field changes were co-spatial with UV emission. In all studied cases the chromospheric brightenings began significantly earlier and ended much later than the photospheric field changes. Thus the authors suggested that the chromosphere must be more responsive to flare energy because more energy from the corona reaches the chromosphere than the photosphere and because the chromosphere is less dense. 

\cite{Cliver12} analyzed photospheric magnetic flux changes and soft X-ray (SXR) emission data and found a sharp change in the unsigned magnetic flux co-temporal with the onset of the flare impulsive phase and the end of the stepwise change co-temporal with the SXR emission peak. Both conclusions, by \cite{Johnstone12} and \cite{Cliver12}, are consistent with the scenario proposed by \cite{Fletcher08} when Alfv\'{e}n waves rapidly transport field changes from the reconnection site in the corona to the photosphere. 

The non-thermal HXR emission in the flaring chromosphere has a strong association with magnetic reconnection in the corona \citep[e.g.,][]{Fletcher02}. HXR emission sources most often appear as kernels at the outer edges of H-alpha/UV ribbons \citep[e.g.,][]{Svestka82}. Recent studies found that both the reconnection and energy release rates are stronger at the ribbon segments accompanied by HXR kernels than at those without HXR sources \citep[e.g.,] [] {Asai04,Temmer07}. Also, the footpoint motions away from the neutral line are considered to be indicative of the reconnection occurring in different heights of arcade magnetic fields with a displacement speed roughly proportional to the rate of reconnection. 

\cite{Cliver12} also looked at the timing of the total flux change and hard X-ray (HXR) emission during three X-class flares and found that the onsets of the stepwise increase in the total flux and the rise in HXR emission are co-temporal. \cite{Yurchyshyn04} investigated the connection between the HXR emission and LOS magnetic field changes associated with X4.8 flare on July 23, 2002 also finding that the start of the sharp rise in HXR emission and the start of the strong stepwise magnetic field changes are co-temporal. \cite{Zharkova05} analyzed the same flare and reported that the magnetic field changes started 6-7 minutes before the HXR emission onset and ended 20 minutes later. Yet, more studies including larger statistical samples, possibly new approaches, and more details on timing of the field changes and flare emissions during different flare phases would be beneficial for a better understanding the relation between the reconnection processes in the corona and the photospheric magnetic filed changes during flares.

In this work, we investigate the spatial and temporal correlation between abrupt permanent changes of magnetic field during five X-class flares observed by Global Oscillation Network Group (GONG) and one X-class flare observed by Helioseismic and Magnetic Imager (HMI) instrument on board Solar Dynamics Observatory (SDO), and the location of HXR emission observed by Reuven Ramaty High Energy Solar Spectroscopic Imager (RHESSI). Our purpose is to explore the likelihood that the photospheric field changes and the chromospheric HXR sources are generally caused by a common physical process in the corona during a flare, and to provide new observational constraints for models of solar flares. We also propose a scenario connecting the photosphere and a reconnection region in the corona that could explain our results.

\section{Observations}

GONG longitudinal magnetic field data were previously used by \cite{Sudol05} to characterize pixel-by-pixel abrupt, stepwise, permanent field changes during 15 X-class flares. \cite{Petrie10} extended this study to a large sample of flares of intensity M5 and above. Following these studies, we analyze $\sim$ 4 hours of GONG line-of-sight magnetic field observations around each flare, and the maps of the field change parameters derived as in \cite{Petrie10}. 

GONG full-disk magnetograms are obtained with 1-minute cadence, 2.5-arcsec pixel size, and 3 G / pixel noise level. The preliminary data reduction procedure is given in detail in \cite{ Petrie10}  and is briefly as follows: (i) the images are remapped to heliographic coordinates $32\arcdeg\times32\arcdeg$ field of view, local to a flare, and tracked; (ii) each image is registered to a reference image formed from the average of the 10 one-minute images immediately preceding the flare. 

The stepwise field changes are modeled as in \cite{Sudol05} by fitting the step-like function:

\begin{equation}
B(t)=a+bt+c \left(1+\frac{2}{\pi} \tan^{-1}(n(t-t_0))\right)
\end{equation}
    
where $t$ is time, $a$ and $b$ model the linear background field evolution, $c$ is the $\frac{1}{2}$-amplitude of the step, $n$ is the inverse time parameter associated with the slope of the step, and $t_0$ is the time at the midpoint of the step. The start time, $t_s$, and the end time, $t_e$, of the field changes are derived from the fit parameters as:  $t_s=t_0-\pi/(2n)$ and $t_e=t_0+\pi/(2n)$.

The magnetic field in each pixel in an active region as a function of time was fitted with Eq. (1) and spatial maps of the fit parameters were created. In the following we refer to these as $c$-, $t_s$-, $t_0$-, and $t_e$-maps.

To avoid spurious fits of Equation 1, the parameter maps include only pixels meeting the following criteria (see \cite{Petrie10} for more details):
(i) exhibiting reasonably sized field changes $\mid 2c \mid < 500$ G;
(ii) with steps of reasonably short duration $< 40$ minutes;
(iii) with the time of the step occurring within 20 minutes of the GOES flare start time. 
We further require that:
(iv) the time series of measurements passed a reduced-$\chi^2$ test;
(v) the background field values were not unreasonably large, $\mid a \mid < 1000$ G. We applied these criteria used by \cite{Petrie10} and found them very helpful in eliminating artifacts in this work. Most pixels with $\mid a \mid > 1000$ G have noisy time series because of the low intensity, and most pixels with $\mid 2c \mid > 500$ G have emission artifacts. It is possible to detect a small minority of good cases with $\mid a \mid > 1000$ G and/or $\mid 2c \mid > 500$ G by eye, but flux integrals are less affected by noise and artifacts with the conditions $\mid a \mid < 1000$ G and $\mid 2c \mid < 500$ G imposed.

The stepwise field changes associated with HXR emission generally have large amplitudes around 100-200 G and above \citep[e.g.,] [] {Fletcher08}. \cite{Sudol05} observed 100-200 G abrupt permanent line-of-sight photospheric magnetic field changes co-spatial with flare ribbons. It has been shown in previous studies \citep[see] [] {Petrie10,Burtseva13} that the amplitude and number of the stepwise field changes are greater for larger flares. Histograms of the field change amplitudes above 50 G for the five flares observed by GONG are shown in Figure 1. For the three stronger flares in this work, there is a good sample of field changes above 100 G, while there are only few of them in the two weaker flares. In order to have a reasonably good sample of the field changes for each flare in our analysis, we exclude pixels with field change amplitude $\mid2c\mid < 50$ G.

\begin{figure}[h]
\centering
\includegraphics[width=.95\textwidth]{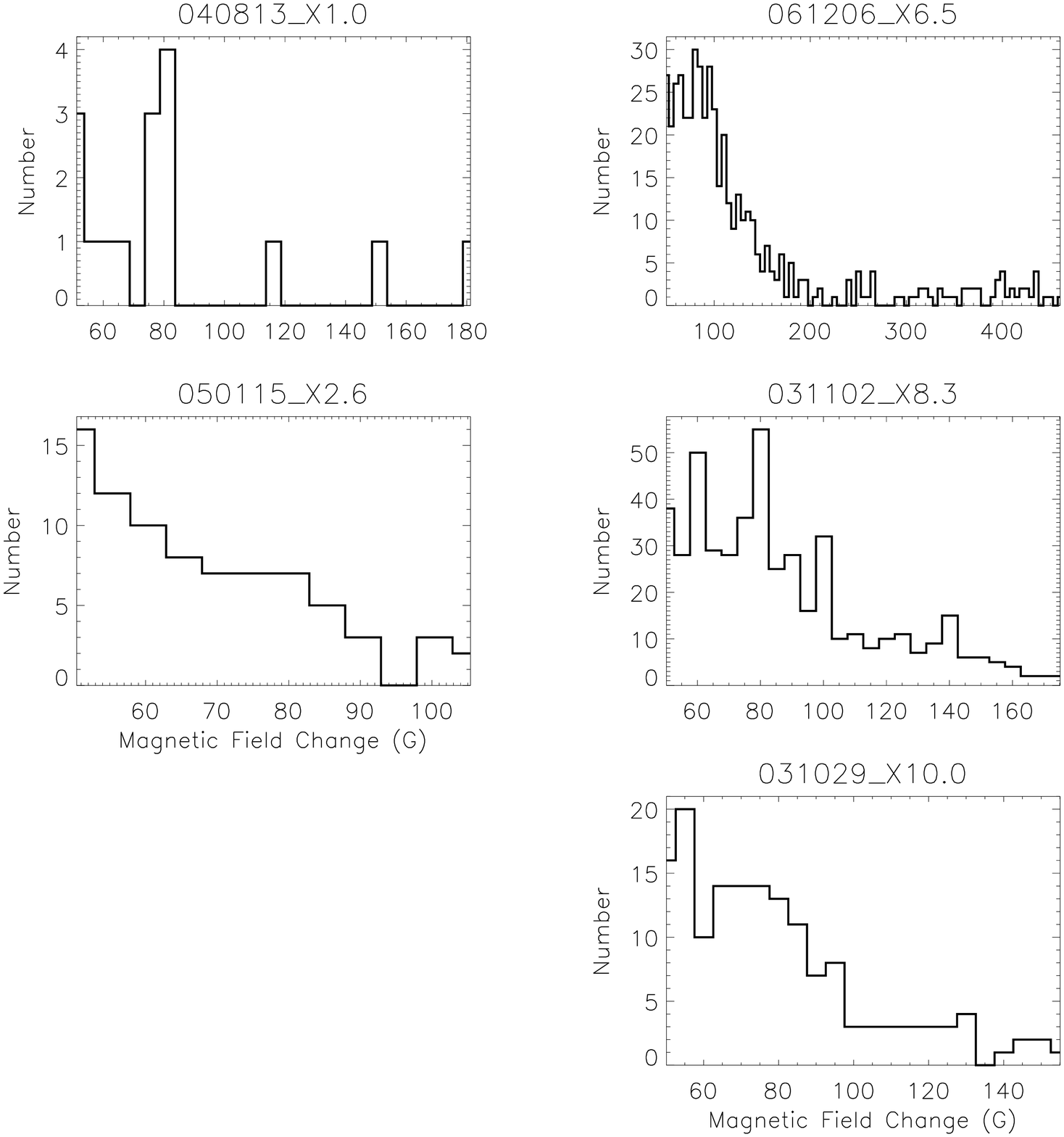}
\caption{Histograms of the field change amplitudes above 50 G for each of the five flares observed by GONG.}
\label{default}
\end{figure}
  
We also use HMI 12-minute vector magnetograms for the X2.1 flare on September 6, 2011 in the form of Active Region Patches (HARPs), maps of $B_r$, $B_{\theta}$ and $B_{\phi}$ with pixel size $0\arcdeg.03$ in heliographic coordinates derived by cylindrical equal-area (CEA) projection \citep[see] [] {Hoeksema14}. A list of analyzed flares with their GOES times, location on the disk and instrument used is given in Table 1.

\begin{table}[h]
\begin{center}
\caption{Date, class, time and location of analyzed flares}
\begin{tabular}{lccccccc}
\tableline\tableline
Date & GOES & GOES Start & GOES Peak & GOES End & Location & NOAA     & Instrument \\
          & Class  & Time (UT)   & Time (UT)     & Time (UT)   &                  & Number &                      \\
\tableline
2003 Oct 29 & X10.0 & 20:37 & 20:49 & 21:01 & S15W02 & 10486 & GONG \\
2003 Nov 2  &  X8.3   & 17:03 & 17:25 & 17:39 & S14W56 & 10486 & GONG \\
2004 Aug 13 & X1.0  & 18:07 & 18:12 & 18:15 & S13W23 & 10656 & GONG \\
2005 Jan 15 & X2.6  & 22:25 & 23:02 & 23:31 & N15W05 & 10720 & GONG \\
2006 Dec 6  & X6.5  & 18:29 & 18:47 & 19:00 & S05E64   & 10930 & GONG \\
2011 Sep 6  & X2.1 &  22:12 & 22:20 & 22:24 & N13W18 & 11283 & HMI \\
\tableline
\end{tabular}
\end{center}
\end{table}   
  
We compare the location and timing of the magnetic field changes and the HXR flare emission derived from RHESSI observations. RHESSI images were synthesized using the CLEAN algorithm \citep[see, e.g.,] [] {Dennis09}. The cadence used varied from 8 to 30 seconds in an energy range from 30 up to 300 keV. The RHESSI images were remapped to the same grid with the corresponding pixel size and tracked to the same minute as the magnetic field data. The centroid position of each HXR source at each moment of time was defined by fitting a two-dimensional elliptical Gaussian above $50\%$ of the local maximum with an accuracy within the pixel size of the data. The total flux of each HXR source at each moment of time was computed within the region of $50\%$ of the maximum flux.  

Also, TRACE 171, 195 and 1600 \AA\ and AIA 171 \AA\ observations were used for context information giving us an idea of the conditions of the upper chromosphere and coronal loop configuration during flares.

\section{Spatial correlation between the field changes and HXR sources}

It has been reported by several authors \citep[e.g.,] [] {Krucker03,Metcalf03,Temmer07}, that HXR sources display a complex motion pattern, where the flare footpoints move along the neutral line in the impulsive phase, and later move away from and perpendicular to the neutral line. \cite{Yang11} see this two-phase motion of the HXR footpoints as a possible indication of of the two-phase magnetic reconnection process, implying the reconnection electric field could have different properties in the impulsive and gradual phases of a flare. In an earlier study \cite{Bogachev05} characterized observed HXR footpoint motions into three types. The first type is a motion away from and perpendicular to the PIL as more flux is reconnected, in agreement with the standard flare model. The second and third types are antiparallel and parallel motions along the PIL, indicating shear relaxation \citep[see, e.g.,] [for supporting observations] {Masuda01,Ji07} and motion of acceleration region in the corona along the separator, respectively. We see all of these motion patterns in the flares we have analyzed. The RHESSI HXR footpoint contours near the emission peak time and temporal evolution of the source centroids positions during the five flares observed by GONG are shown in Figures 2 and 3, and those during the flare observed by HMI are shown in Figure 4. 

     In the three stronger flares in our study, observed by GONG (see Figure 3), we see antiparallel motion of the main HXR sources (labeled S1 and S2) along the PIL and then away from the PIL after the HXR flux has reached its maximum. Besides the two main sources, a third HXR source, labeled S3, appears in two of the flares, the X8.3 and X10.0. It is seen for a short time only in the beginning of the HXR emission and does not show a clear pattern in its motion. It also could be a part of the S2 HXR footpoint, but appears as a separate source at the $50\%$ level of the maximum HXR flux. The S1 footpoint in the X6.5 flare is much weaker than the S2 footpoint. It dims very quickly and does not move as much as the S2 source moves. This asymmetry in the brightness of the two footpoints could be indicative of an asymmetric loop. Also the HXR footpoint in the stronger magnetic field appears to move more slowly than in the weaker magnetic field \citep[see, e.g.,] [] {Jing08,Yang11}. The slower HXR motion in stronger compared to weaker field is consistent with an approximately steady reconnection rate. 
     
     In one of the weaker flares, observed by GONG (see Figure 2, bottom), the X2.6 flare, we notice parallel motion of the S1 and S2 HXR sources along the PIL in the impulsive phase of the flare, and an asymmetric motion after the strongest HXR peak: continuous motion of S1 along the PIL and motion of S2 away from the PIL. The S3 and S4 HXR sources, appearing in the eastern part of the region, also show an asymmetric motion. The S3 footpoint moves along the PIL towards the south-east and then turns around and moves towards the north-west in the direction opposite to the motion of the S1 footpoint. The S4 source moves parallel to the S3 and then away from the PIL. \cite{Liu10} have also analyzed the asymmetric motion of the HXR footpoints in this flare and interpreted it as an asymmetric eruption and thus magnetic reconnection progressing along the PIL. 
     
     In the other weaker flare (see Figure 2, top), the X1.0 flare, there is only one detectable HXR source, moving parallel to the PIL and apparently above the PIL. In the movies of the TRACE 171 \AA\ observations (see Section 4), where HXR sources were over-plotted, it also looks like a motion of a loop top rather than a footpoint motion, however it is hard to say with certainty. Loop-top sources are most easily observed in limb flares whose loop footpoints are occulted by the limb. Footpoint sources are stronger than their coronal counterparts during the impulsive phase of the flare. After the impulsive phase the footpoints fade and the loop-top becomes brighter, nevertheless, coronal sources above 30 keV are rarely observed \citep[e.g.,] [] {Tomczak07}. 
          
     In the X2.1 flare observed by HMI two footpoints move along the PIL in the same direction (see Figure 4) and seem to start turning away from the PIL at the last data point. However, it is hard to say what happened beyond the time frame of the HXR data set we have for this flare. The spatial location of the two footpoints coincides with the location of the strongest magnetic field changes in both radial, $B_r$, and horizontal, $B_h$, vector field components and corresponds to the strong tilt angle change.    

In all of the analyzed flares, the early RHESSI emission corresponds well to locations of strong field changes. The strong field changes occur predominantly in the regions of strong magnetic field near the PIL. The later RHESSI emission does not correspond to significant field changes as the footpoints are moving away from the PIL.   

\begin{figure}[t]
\centering
\includegraphics[width=.95\textwidth]{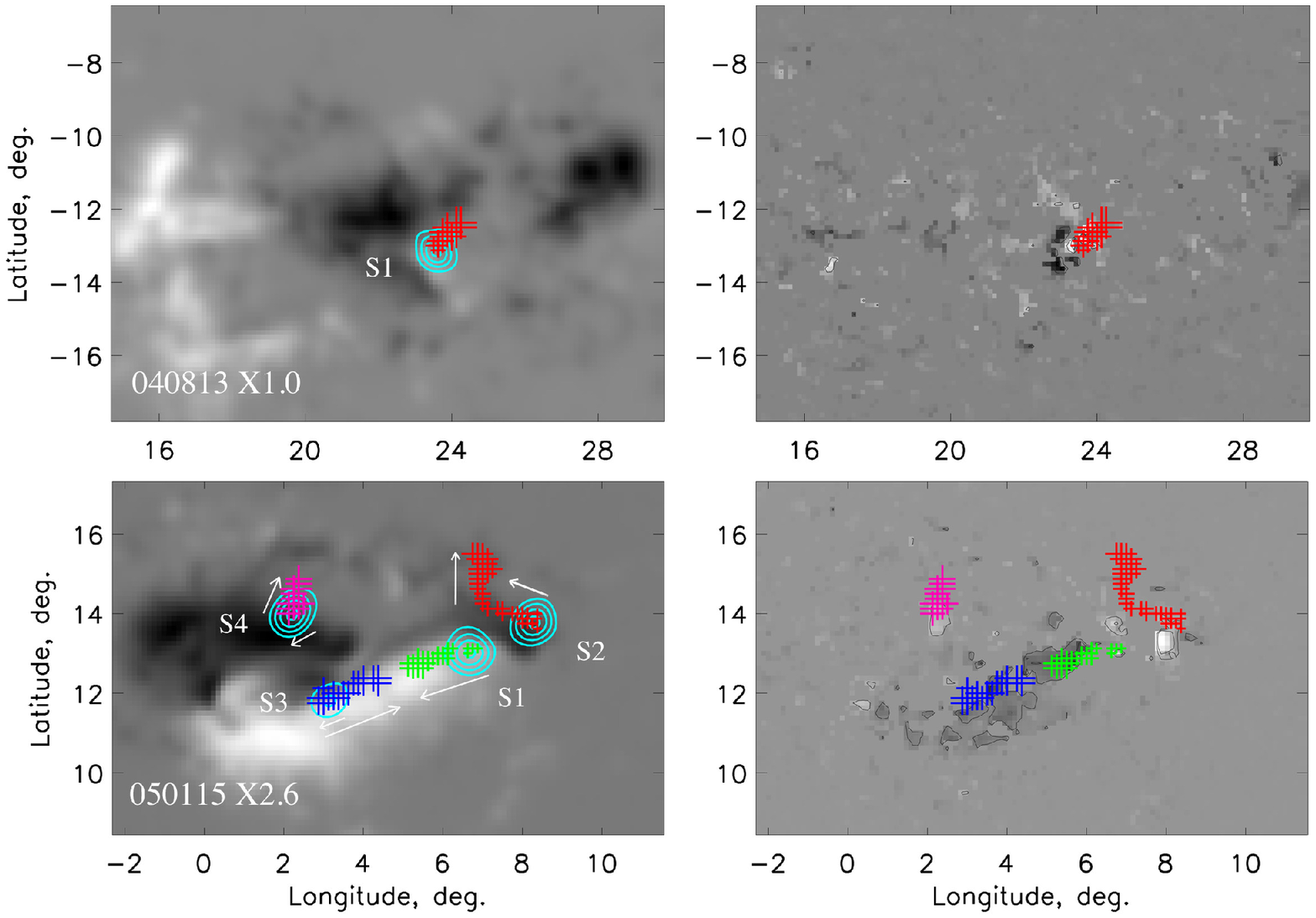}
\caption{LOS magnetogram of the flared active region (left), cropped to the area of the strongest field changes, represented by the fit parameter c (right) for the X1.0 and X2.6 flares, the two weaker flares in our analysis observed by GONG. Location of the strong field changes is indicated by gray contours at 50 and 100 Gauss of the field change amplitude. RHESSI 30-70 keV and 40-100 keV HXR footpoint contours for the two flares, respectively, at levels from $35\%$ to $95\%$ of the maximum counts at around the emission peak time are superposed in light blue. The temporal evolution of the source centroids is shown with plus symbols, which increasing size represents progressing times during flares and different color denotes different sources. In addition to that, direction of the HXR footpoints motion for the X2.6 flare is indicated by arrows for clarity purposes. HXR sources are labeled by 'S' letter with a number.}
\label{default}
\end{figure}

\begin{figure}[h]
\centering
\includegraphics[width=.95\textwidth]{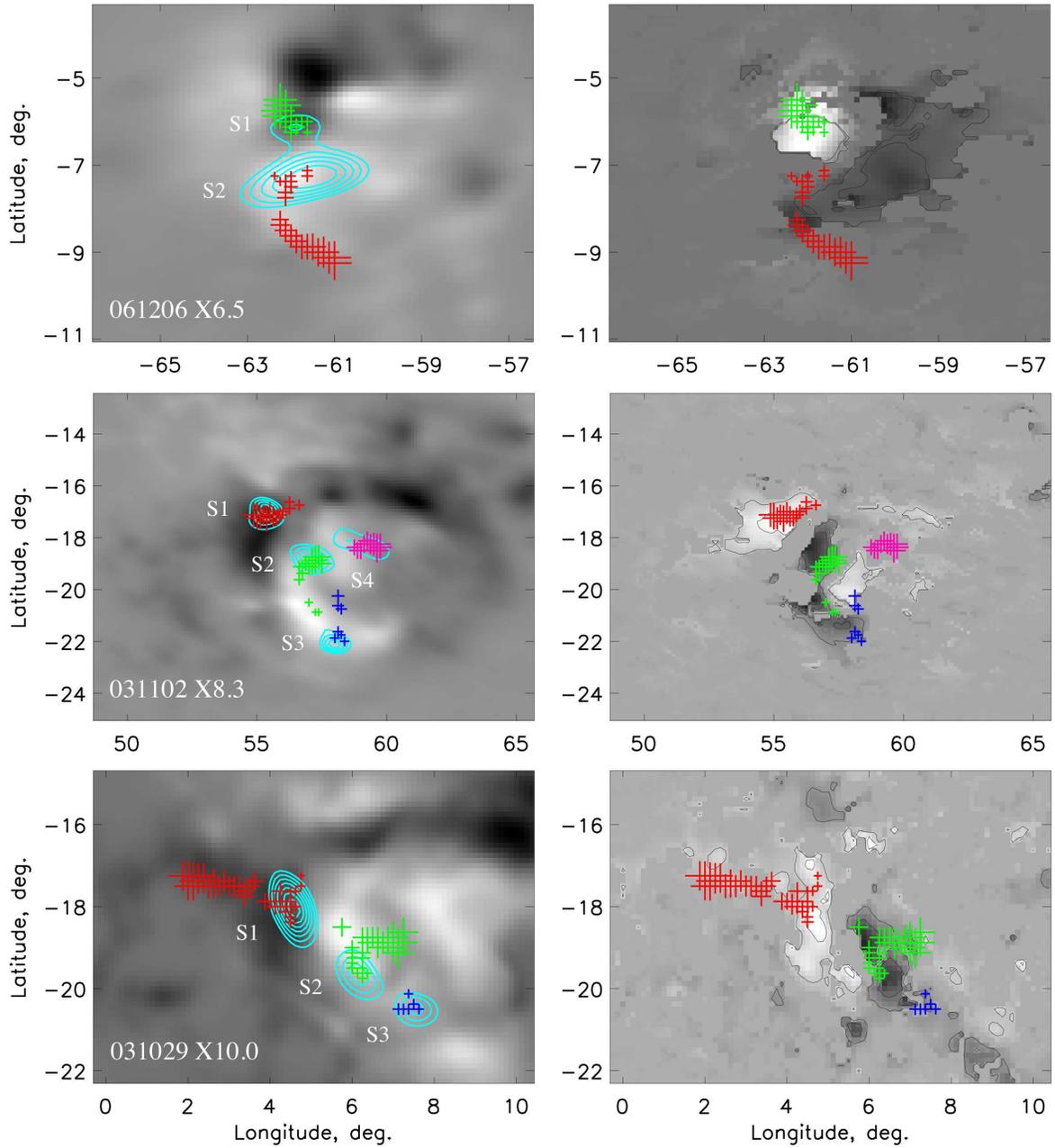}
\caption{The same as Figure 2, but for the X6.5, X8.3 and X10.0 flares, the three stronger flares in our analysis observed by GONG. RHESSI 50-300 keV, 50-300 keV and 50-100 keV HXR footpoint contours for the three flares, respectively, superposed in light blue are at levels from $50\%$ to $90\%$ of the maximum counts at around the emission peak time.}
\label{default}
\end{figure}

\begin{figure}[h]
\centering
\includegraphics[width=.95\textwidth]{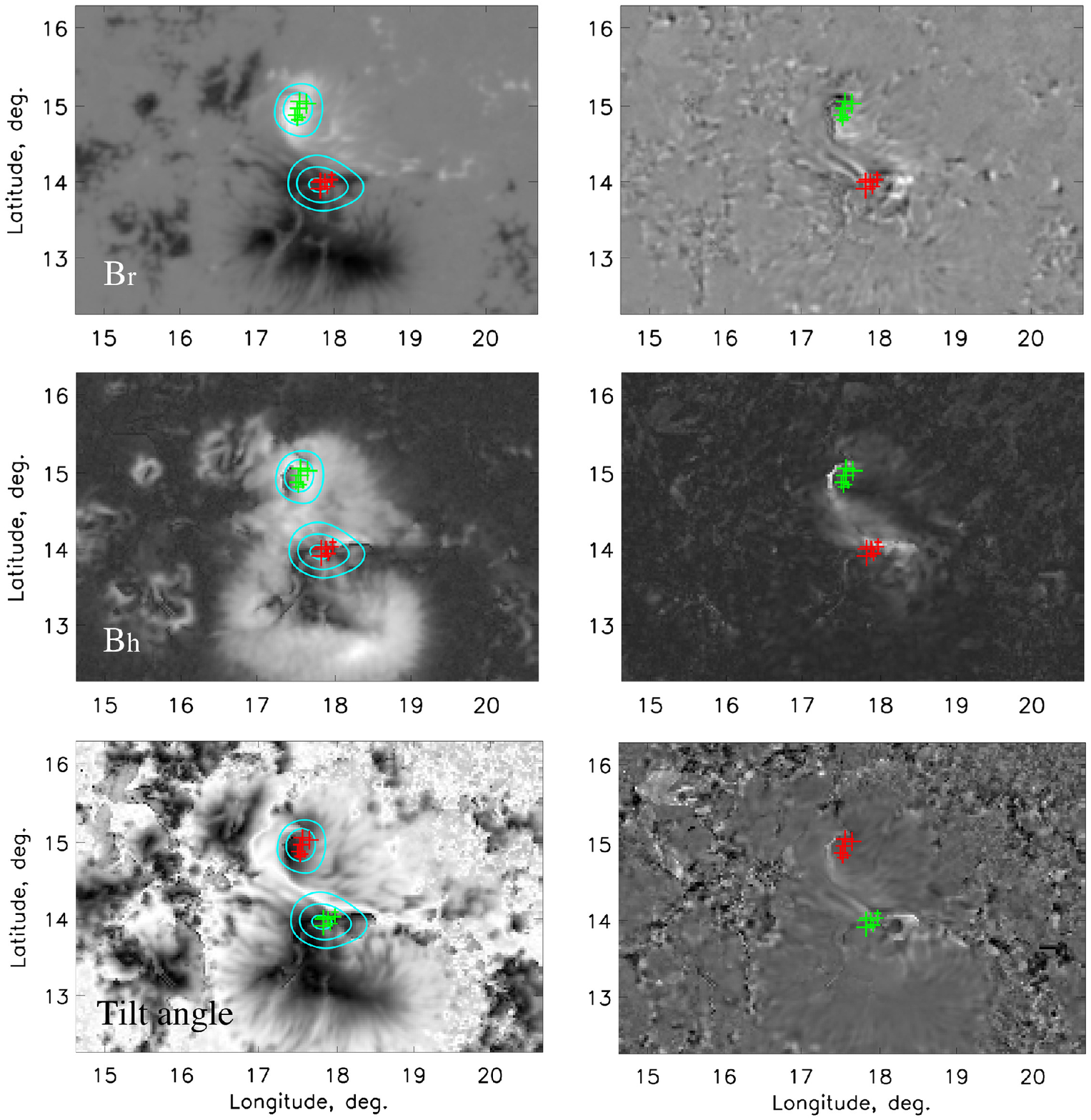}
\caption{Vector magnetic field of the flared active region (left) and the vector magnetic field changes (right) during the X2.1 flare observed by HMI. From top to bottom: the vertical field component, the horizontal field component and the field tilt angle. RHESSI 50-160 keV HXR footpoint contours at levels from $50\%$ to $95\%$ of the maximum counts at around the emission peak time are superposed in light blue. The temporal evolution of the source centroids is shown with plus symbols, which increasing size represents progressing times during flares and different color denotes different sources.}
\label{default}
\end{figure}

\section{Comparison of HXR with UV and EUV observations from TRACE and AIA}

In order to visualize the location and evolution of the HXR footpoints during a flare with respect to the coronal loop configuration, we over-plot the HXR sources on top of TRACE 171, 195 and 1600 \AA\ images (subject to availability of the particular TRACE data) and run the movies constructed from this data. The comparison suggests that the HXR footpoints seem to be confined to short, bright loops. We see that, in agreement with previous studies \citep[e.g.,] [] {Masuda01,Fletcher02,Cheng12}, the HXR sources are well correlated in space and time with bright kernels of the UV and extreme UV (EUV) emission. GONG intensity and TRACE WL observations (subject to availability) show the HXR footpoints moving roughly along the umbral-penumbral boundaries in the sunspots, consistent with previous studies \citep[e.g.,] [] {Temmer07}.

\cite{Johnstone12} found a pattern linking 1600 \AA\ brightenings and longitudinal field changes: the brightenings occurred wherever field changes took place but not vice versa. In terms of the temporal correlation between field changes and flares, the authors found that the UV brightenings began before and ended after the field changes. We investigate the temporal relationship between the HXR emission and magnetic field changes during the flare in the next Section.

\section{Temporal correlation between field changes and HXR sources}

In order to determine the temporal relationship between the magnetic field changes and HXR emission during flares, we perform the following analysis. First, we analyze the step-function fit parameter maps (the field change amplitude, $c$-, start time, $t_s$-, midpoint time, $t_0$-, and end time, $t_e$-maps), derived from GONG magnetic field data using Eq. (1), along with temporal evolution of the HXR flare emission. Second, we examine the temporal and spatial correlation of the field change and the HXR signal by tracking the field changes directly from magnetograms with the HXR footprint contour masks at each moment of time. We also analyze the timing of the total flux change of the whole flaring region and at the locations of the strongest field changes next to the PIL. 

\subsection{Analysis of the step-function fit parameter maps} 

Each of the $t_s$-, $t_0$- and $t_e$-maps provides information about the field change timing in each pixel where a strong stepwise change occurred during a flare. We construct a histogram of the field change over time by integrating amplitudes of the field changes from the $c$-maps over one-minute intervals around each flare for each of the start, midpoint, and end time from $t_s$-, $t_0$- and $t_e$-map, respectively. Figure 5 shows the computed total magnetic flux change as well as total flux of the HXR signal as functions of time. Most of the field changes start before or around the start time of the detectable HXR flare emission, and they end at about the same time or later than the detectable HXR signal. The three strongest flares, the X6.5, X8.3 and X10.0, show a peak in the total magnetic flux change $\sim$ 1.5, 3 and 4 minutes, respectively, earlier than the peak in the HXR signal. The maximum in the flux change appears at or around the same minute in all three histograms, for the start, peak and end times, indicating that the largest flux changes take a short time to occur. More than $54 \%$ of the field changes are complete within a minute, and more than $75 \%$ of them are complete within 3 minutes. These fast field changes do not occur at other times. 

The Levenberg-Marquardt least-squares function fitting method \citep[][]{Press07} was used to fit Eq. (1) to the magnetogram pixel time series. Besides values for the parameters, this algorithm also produces values for the associated variances, which we can use to estimate the errors associated with the parameter values. The estimated midpoint-time $t_0$ of a stepwise field change has a smaller error than the estimated start and end times $t_s$ and $t_e$, as the latter ones strongly depend on the duration of the step: the longer the step duration, the larger the uncertainty of the $t_s$ and $t_e$ \citep[see][for details]{Petrie10}. The mean uncertainty in $t_0$ in our analysis is $\sim 0.28$ minutes, which is smaller than the time differences between the peak in total flux change and the peak in the HXR signal found for the three strongest flares in this work. 

The two weaker flares do not  seem to have a strong peak in the flux change at a certain time and the field changes do not show clear temporal relation to the HXR signal. For these flares, only very few pixels had field change amplitudes around or above 100G, and in general the number of field changes that passed the rejection criteria and were larger than 50G is less than half as many as for the three strongest flares (see Figure 1). The total flux changes for the  weaker flares are at least 4 times smaller than the total flux changes for the stronger flares.  

\begin{figure}[h]
\begin{center}
\includegraphics[width=.95\textwidth]{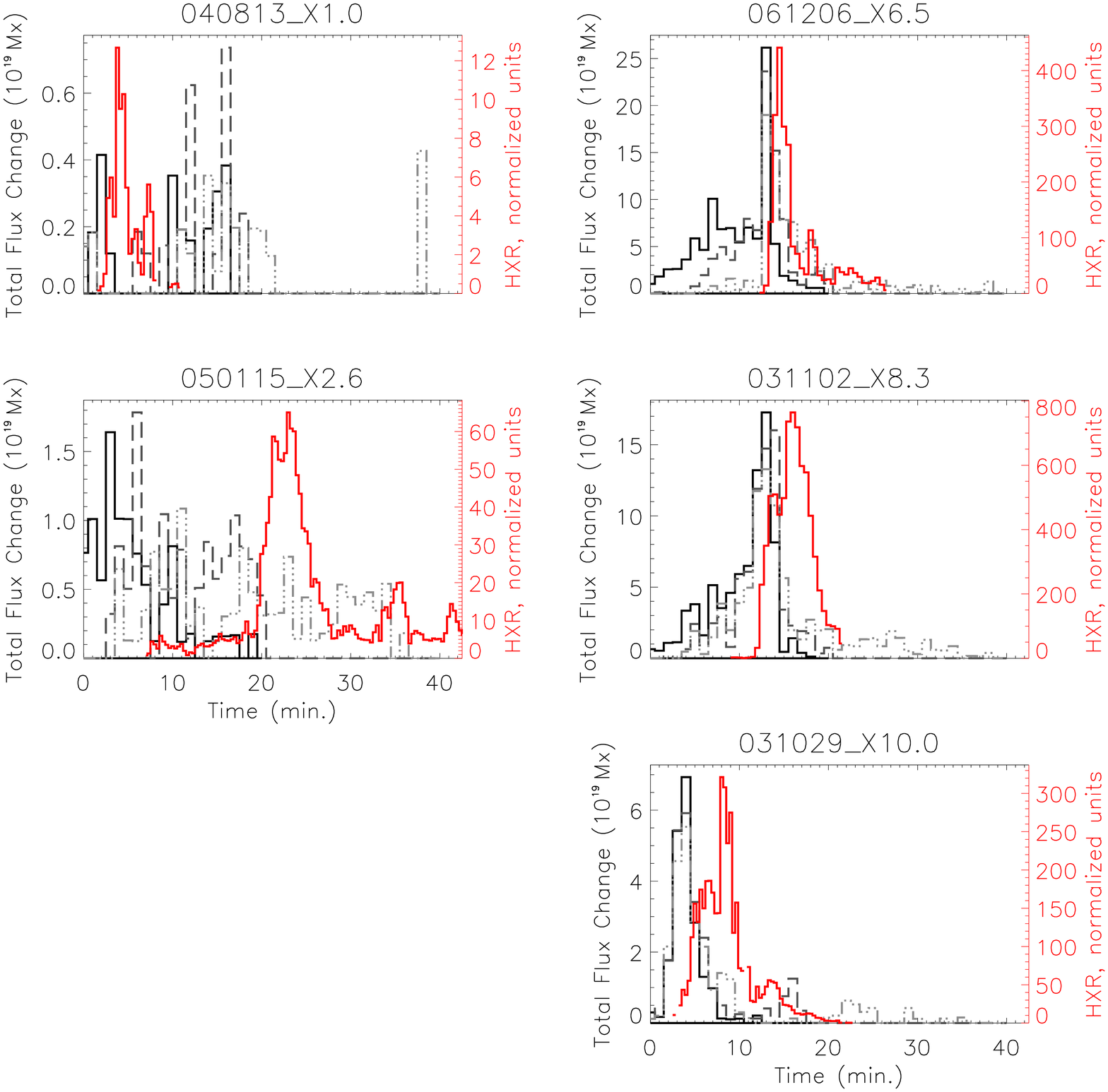}
\caption{Total flux change as a function of time during each of the five flares observed by GONG. Time axis is in minutes relatively the GOES flare start time. Black solid, dark-grey dashed and light-grey dash-dotted lines correspond to the start, midpoint and end times, respectively. Total HXR flux is over plotted in red.}
\label{default}
\end{center}
\end{figure}
 
 To understand why the stronger X-class flares show a strong peak in the flux change, which does not seem to be the case for the weaker flares, we compute the total background magnetic flux in the pixels, contributing to the total flux change in Figure 5, at each minute. The background magnetic field is derived from a reference image composed of ten remapped GONG magnetograms immediately preceding the flare. The total unsigned background field, shown in Figure 6, is highly correlated with the total unsigned flux change that was also found in previous studies \citep[see][]{Petrie10,Burtseva13}.    
 
 \begin{figure}[h]
\begin{center}
\includegraphics[width=.95\textwidth]{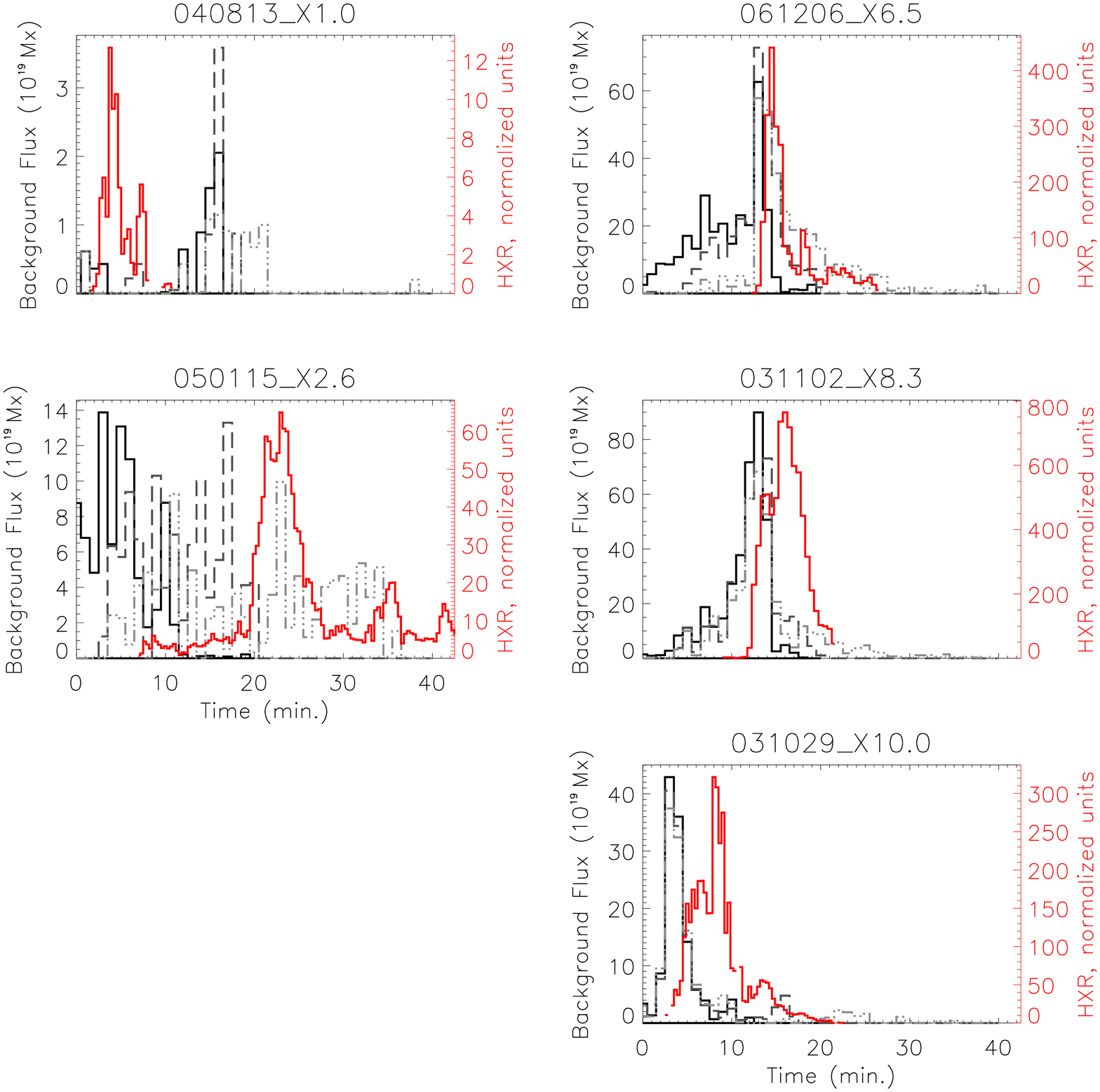}
\caption{Total background flux as a function of time during each of the five flares observed by GONG. Time axis is in minutes relatively the GOES flare start time. The background flux is obtained for the same pixels used to compute flux change shown in Figure 5.}
\label{default}
\end{center}
\end{figure}
 
For the larger statistical sample of 77 flares (39 M-class and 38 X-class) studied by \cite{Petrie10}, we now analyze the total flux change as a function of time, and compute their histograms in the same way as we did for Figure 5. Visually we find that all of the flares above $\sim$ class X5 show a strong peak in their flux change over a short timespan. Some of the weaker flares also show a peak but the peak is lower and not as sharp relative to the amplitudes of the flux changes at other times. 

To generalize the result to the larger set of 77 flares, we compute the relative difference between the maximum peak value and median value of the total flux change during flare time, where the median is the reference. Some of the flare data sets include very few pixels whose flux changed in a stepwise manner during the flare. Fewer pixels recorded stepwise field changes in general during weaker flares \citep[see][]{Petrie10,Burtseva13}. We ignore those cases where the number of pixels with significant flux change during a flare was $< 10$. The relative difference as a function of the GOES flare class is shown in Figure 7. A linear trend is clearly seen on the plot. The Pearson linear correlation coefficients for the start, midpoint and end times are 0.65, 0.51 and 0.52 and their $P$-values, the probability that the observed correlation occurs by chance, computed using the t-test, are $4.3\times10^{-5}$, $3.1\times10^{-3}$ and $1.5\times10^{-3}$, respectively, revealing a moderate correlation between the relative peak height in the flux change and GOES flare class. This confirms that the stronger flares, in comparison with the weaker flares in our sample, in general, show a stronger peak in the flux change.     

 \begin{figure}[h]
\begin{center}
\includegraphics[width=.95\textwidth]{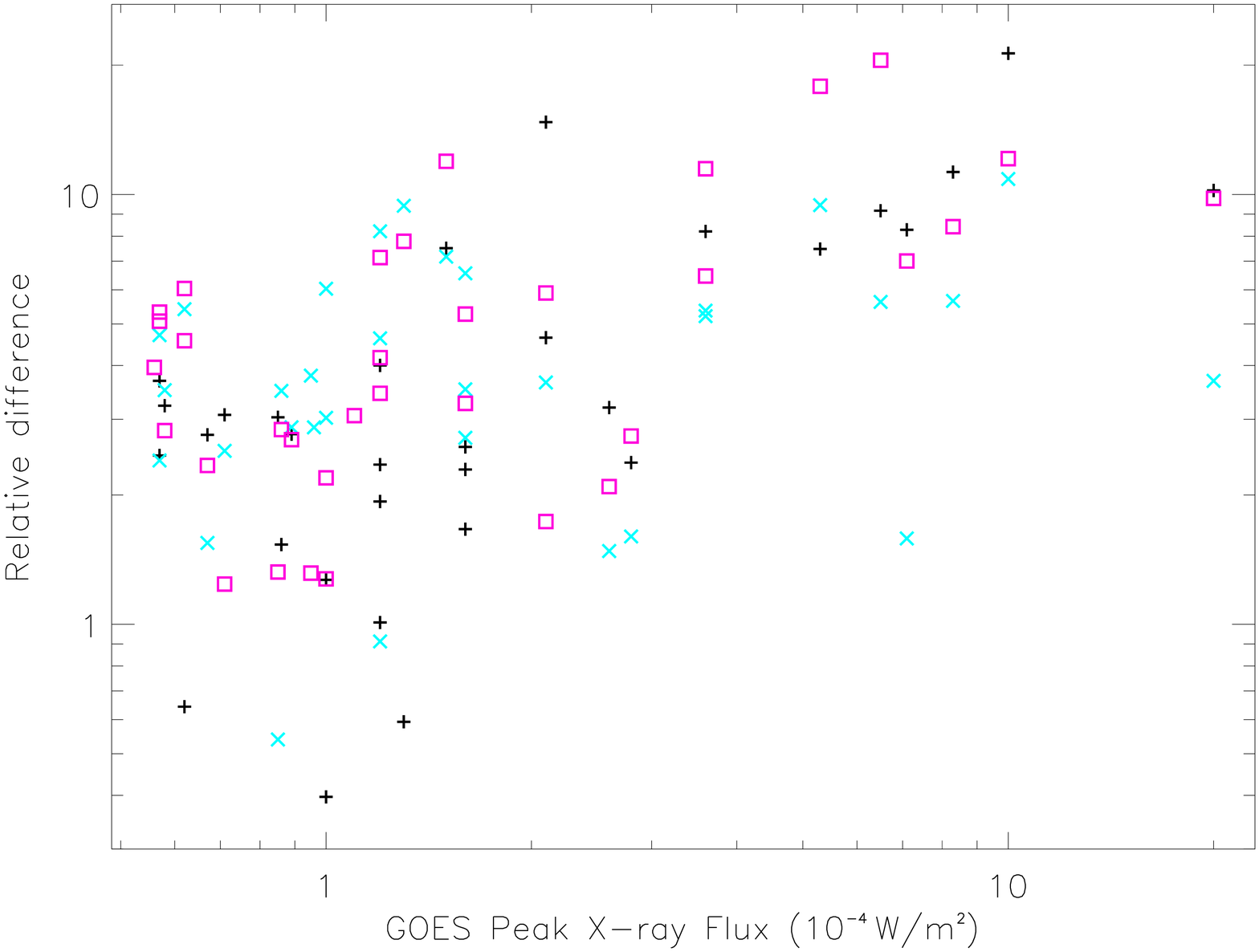}
\caption{Relative difference between the maximum peak value and median value of the total flux change during flare time as a function of the GOES flare class. Black pluses, light-blue crosses and magenta squares correspond to the start, midpoint and end times, respectively.}
\label{default}
\end{center}
\end{figure}

\subsection{Tracking field changes in magnetograms} 

We also analyze the correlation between the field changes and HXR flare emission directly from GONG magnetic field data, tracking the field changes with RHESSI footpoint contours at 50$\%$ of the emission peak. We sum the magnetic field inside the footpoint contours at each footpoint location at each moment of time. As a result we have  temporal profiles of the total flux at each location of the HXR sources during each flare. An example of the flux profiles for one of the HXR sources of the X10.0 flare is shown in Figure 8. The profiles are smoothed with a Gaussian kernel. The width of the Gaussian window varies from 2 to 3 minutes for different flares depending on the noise level in their magnetic flux profiles. We define the midpoint time of the any profile showing a clear step as a maximum gradient of the step. Since often these profiles have a spike (the magnetic transient, see Figure 8 around 20:45 UT on the x-axes), we also define the time of the spike. 

The midpoint times of the field changes as a function of timing of the HXR source are shown in Figure 9. We find that most of the field changes occurred earlier than the HXR signal was detected in the region. Some of the field changes propagate with speed close to that of the HXR footpoint, but at a later phase of the flare, often after the HXR signal reaches its maximum and the footpoint has begun to move away from the PIL. Thus, the field changes follow the same trajectory as the HXR footpoint, but at an earlier time. Most of these field changes are decreases. On the other hand, there are field changes that occur at later times, but at the locations of earlier HXR sources close to the PIL; these tend to be increases: to see this, refer to the timing for the HXR source S2 of the X6.5 flare and for the HXR source S1 of the X2.6 flare in Figure 9. 

The X1.0 flare had only one well detected HXR source. As mentioned in Section 3, this HXR footpoint moved along the PIL, about half of it on the negative polarity side and about half on the positive polarity side. It appears that the negative polarity unsigned total flux decreases and the positive polarity unsigned total flux increases during the flare, both showing a clear abrupt step. The average unsigned total flux over the whole footpoint does not show a clear step. Thus the times for this event were defined separately for the negative, S -1, and positive, S +1, polarity part of the HXR footpoint.  As seen in Figure 9, both the field changes propagate with nearly the same or slightly slower speed as that of the footpoint, but most of them, both the increases and the decreases, occured at a later time, after the footpoint has passed through the region.       

\cite{Sudol05} analyzed field changes during the X2.6 flare on December 11, 2001 and found that the field changes propagate with the H$\alpha$ flare ribbon at a similar speed. They do not mention whether the field change occurred earlier or later than the flare ribbon. They saw more cases of propagation of the field changes across the active region, but in all of the cases the field changes were too fast and restricted to a small area to be accurately tracked. \cite{Petrie10} also notice a propagation of the field changes starting from near the PIL across the region on the $t_0$-map of the X6.5 flare on December 6, 2006.  

The so-called Ómagnetic transientÓ (MT) or Ómagnetic anomalyÓ, significant reversible magnetic field change (as opposed to stepwise permanent field changes) that appears as a ÓspikeÓ in some of our temporal flux profiles during the impulsive phase of a flare, was observed by many authors \citep[e.g.,] [] {Kosovichev01,Qiu03,Maurya12,Harker13}. \cite{Kosovichev01} and \cite{Harker13} suggest that this phenomenon could be a real change in magnetic field, while \cite{Qiu03} and \cite{Maurya12} conclude that it is a result of a line-profile change during flare. The subject is still under debate. In this paper we do not attempt to analyze the nature of the MT, but we describe our observations. We measure the time of the flux minimum in the MT and plot it as a function of HXR time in Figure 10. The MTs associated with HXR sources S1 and S2 in the X10.0 flare, and source S2 in the X6.5 flares are co-spatial and co-temporal with the HXR footpoint in the later phase of the flares, after the HXR emission maximum, when the footpoints are moving away from the PIL. In contrast, the MT associated with source S2 in the X2.6 flare propagates with the footpoint earlier in the impulsive phase of the flare. In case of the HXR source S1, S3 and S4 in the X2.6 flare, and all HXR sources in the X1.0 and X8.3 flares, the MTs are not found to be propagating with the footpoints. They occur along the footpoint trajectory approximately at the time of the HXR emission peak. We also note a transient increase in the magnetic field in two flares, the X1.0 flare in the S -1 location and the X6.5 fare in the S2 later location that appears ahead of the downward spike time.

\begin{figure}[h]
\begin{center}
\includegraphics[width=.95\textwidth]{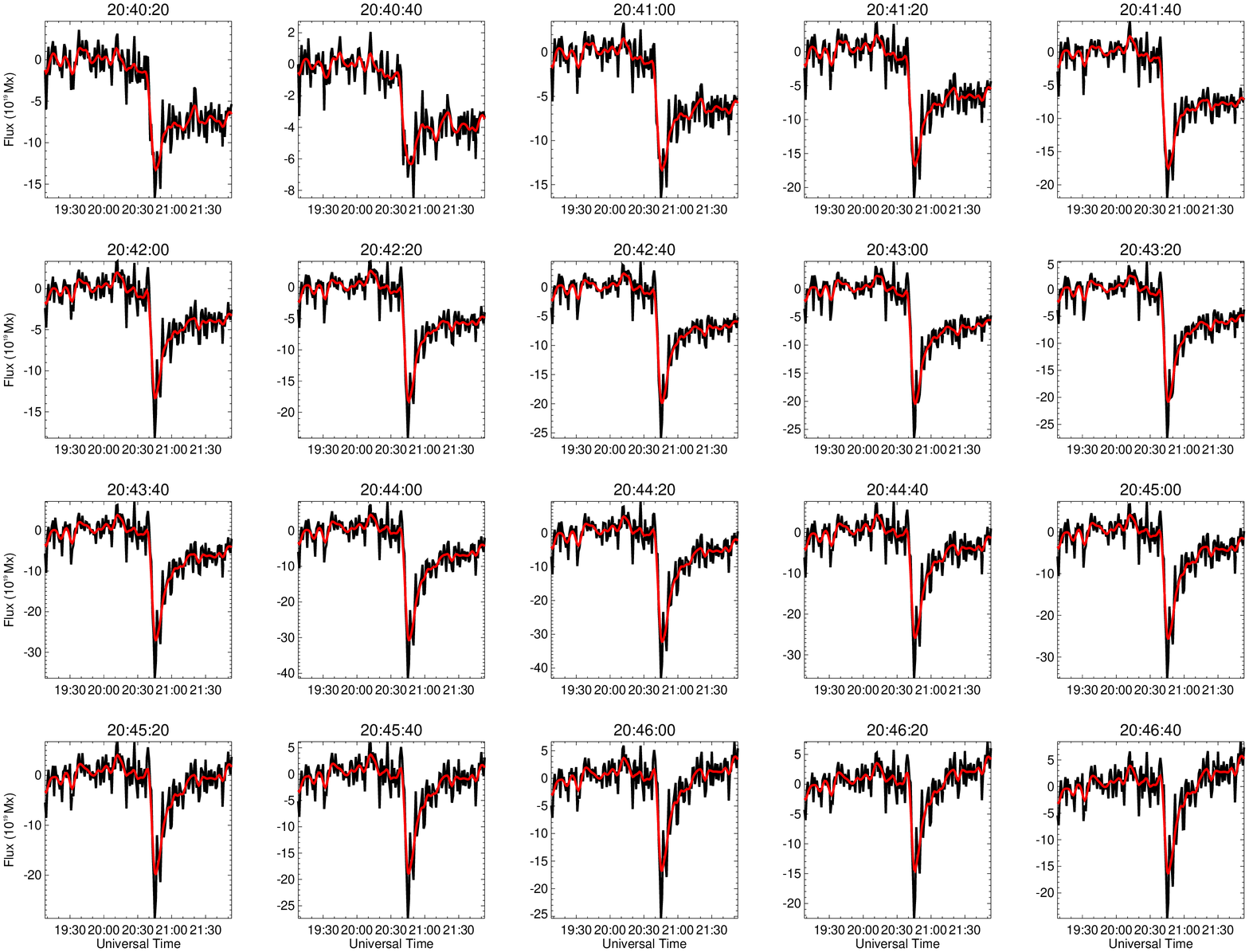}
\caption{Magnetic flux as a function of time (black line) at each location of the HXR source S1, for first 6 min. 20 s. of the HXR signal data, during X10.0 flare on October 29, 2003. The magnetic flux time series smoothed with Gaussian kernel is over plotted in red. X-labels show universal time of the magnetic field data, while the titles represent universal time of the HXR signal.}
\label{default}
\end{center}
\end{figure}

\begin{figure}[h]
\begin{center}
\includegraphics[width=.95\textwidth]{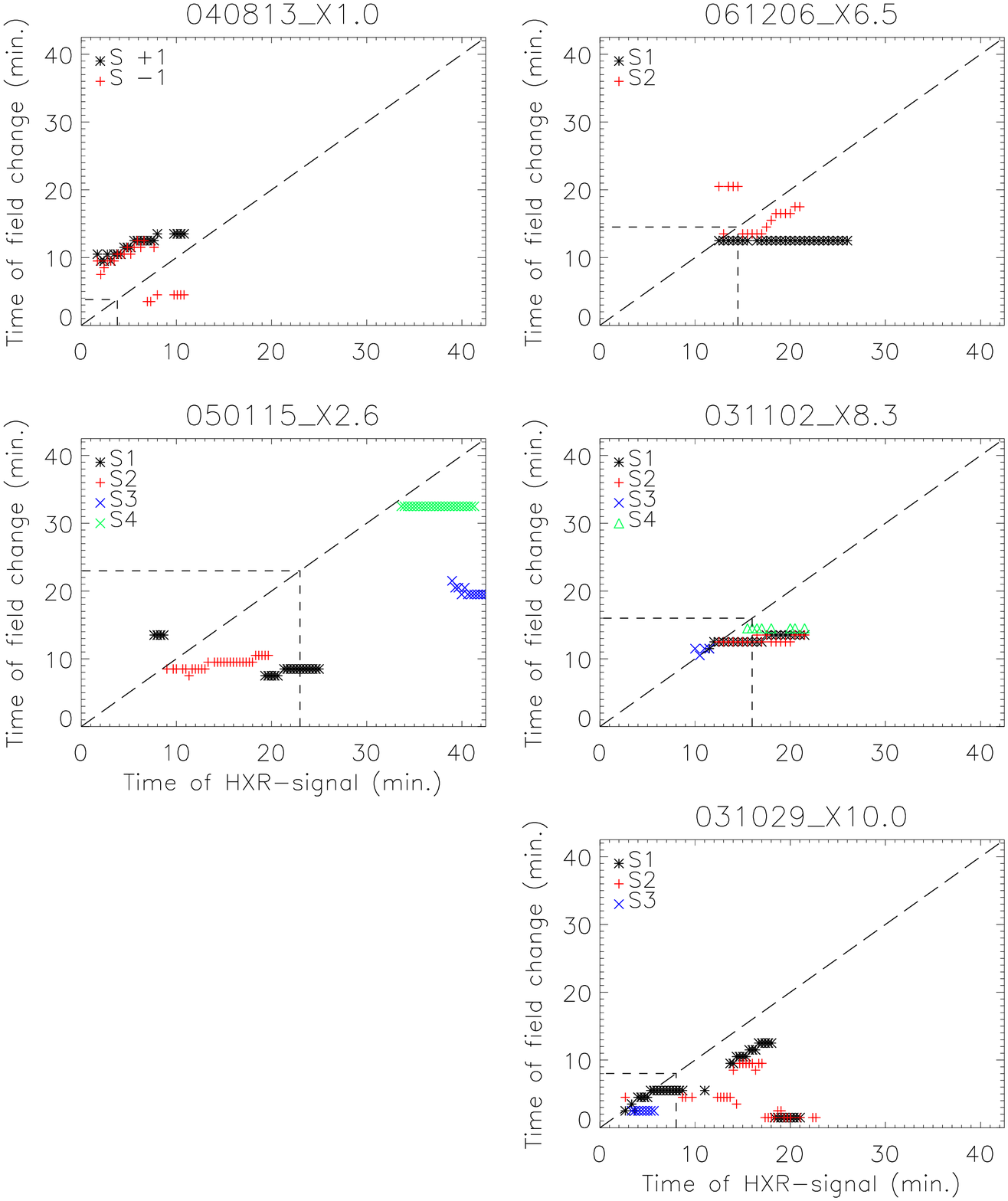}
\caption{Midpoint time of the field change as a function of timing of the HXR signal in each location of the HXR flare sources. Time axes are in minutes relatively the GOES flare start time. Dashed lines creating a rectangle in each panel indicate HXR peak time for the flare on the both axes.}
\label{default}
\end{center}
\end{figure}

\begin{figure}[h]
\begin{center}
\includegraphics[width=.95\textwidth]{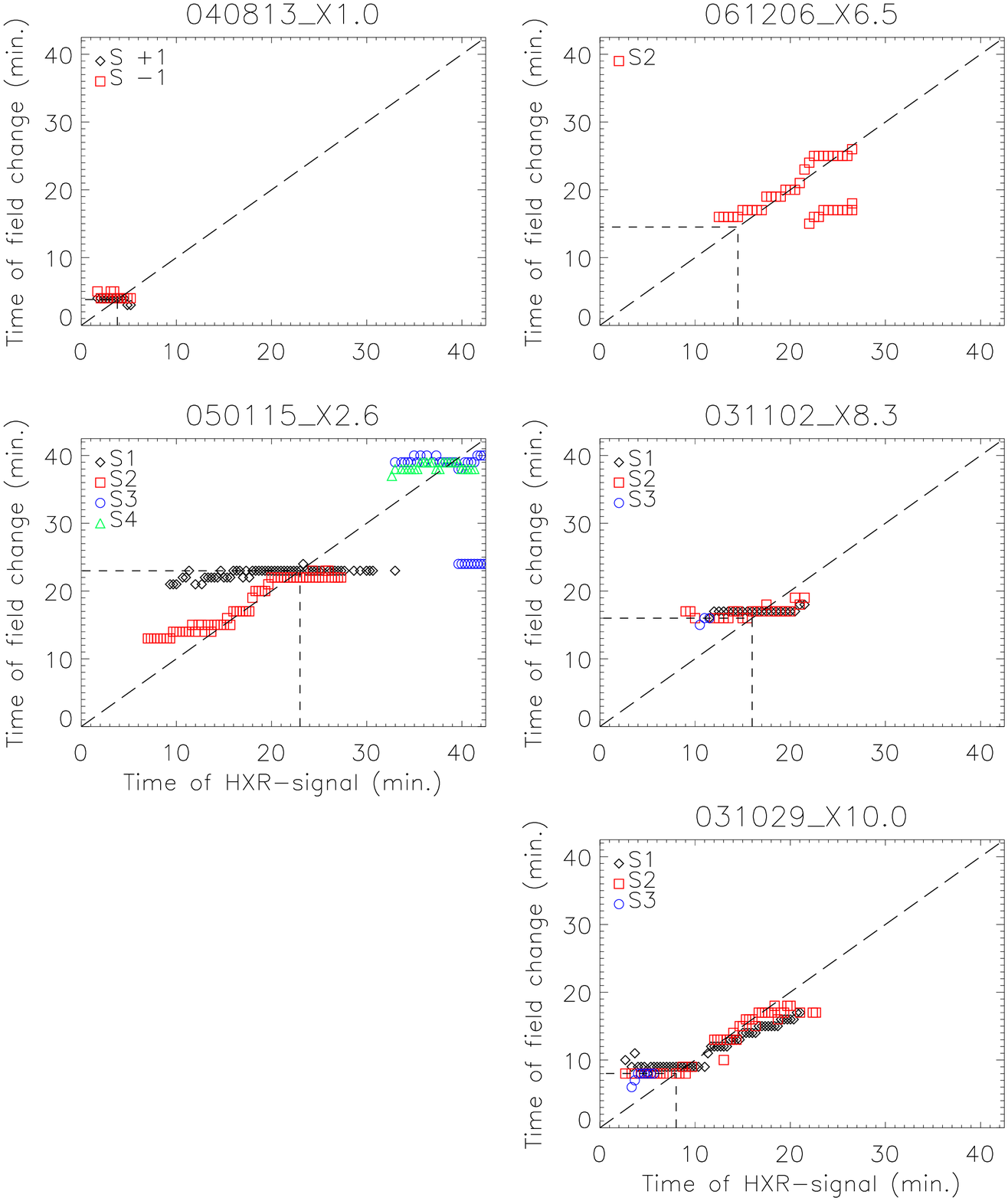}
\caption{Time of the magnetic transient as a function of timing of the HXR signal in each location of the HXR flare sources. Time axes are in minutes relatively the GOES flare start time. Dashed lines creating a rectangle in each panel indicate HXR peak time for the flare on the both axes.}
\label{default}
\end{center}
\end{figure}

\subsection{Flux change in the whole region and at the PIL} 
We compute the total unsigned flux over a rectangular domain including the whole flaring active region cropped to the area of the strongest field changes. For the five flares observed by GONG areas of magnetic field integration correspond to the regions shown in Figures 2 and 3. For the flare observed by HMI this area refers to a smaller, 2 degrees in latitude by 1 degree in longitude, rectangular region around center of the region shown in Figure 4. The magnetic fluxes in all of the flaring regions show a stepwise change associated with the flare (Figure 11). In some of the flares the step is clear and abrupt, whereas in others it is not as strong and obvious on top of the noise or sharp short-time transient field change during the flare. We see that the step-wise field change is co-temporal with the the total HXR flux for most of the flares. The field change starts at around the time when emission in the HXR sources is first detected, and it ends at around the same time as the HXR sources. It is consistent with the results of \cite{Cliver12} who averaged all pixels' profiles in a flaring region showing strong stepwise change and reported that the onsets of the stepwise rise in the total flux and the rise in HXR emission are co-temporal. 

\cite{Petrie10} and \cite{Burtseva13} found that significantly more magnetic flux decreases than increases occurred during the flares, consistent with a model of collapsing loop structure for flares. Consequently, integrated magnetic field over a flared active region should most likely show a decreasing change in the field. In this work the total unsigned fluxes in all flares, observed by GONG, decrease. In \cite{Cliver12} unsigned fluxes increase by construction, as all pixels' profiles were arranged so that they all changed in the same sense, thus preventing the individual pixels' steps from canceling each other. 

The flare observed by HMI shows an increase in total unsigned flux for both vertical, $B_r$, and horizontal, $B_h$, field components, and the average tilt angle. The change in the Bh is much stronger than in the $B_r$ as was found also by \cite{Petrie12}. The $B_r$ component shows the magnetic transient in its profile, while $B_h$ does not. The 12-minute vector data seems to resolve the overall flux change over the rectangular 2$\times$1 degree area adequately - the 12-minute cadence is smaller than the duration of the flux change, and the recorded HXR signal with duration $\sim$ 2 minutes appears more or less in the middle of the step. The tilt angle changed not significantly as a result of this flare, but the tilt relative to the associated potential field did change significantly, consistent with a release of magnetic stress during the flare \citep[see][]{Petrie12}. However, for more detailed comparison with HXR one has to look at another flare where a longer HXR emission time series is available.

\begin{figure}[h]
\begin{center}
\includegraphics[width=.95\textwidth]{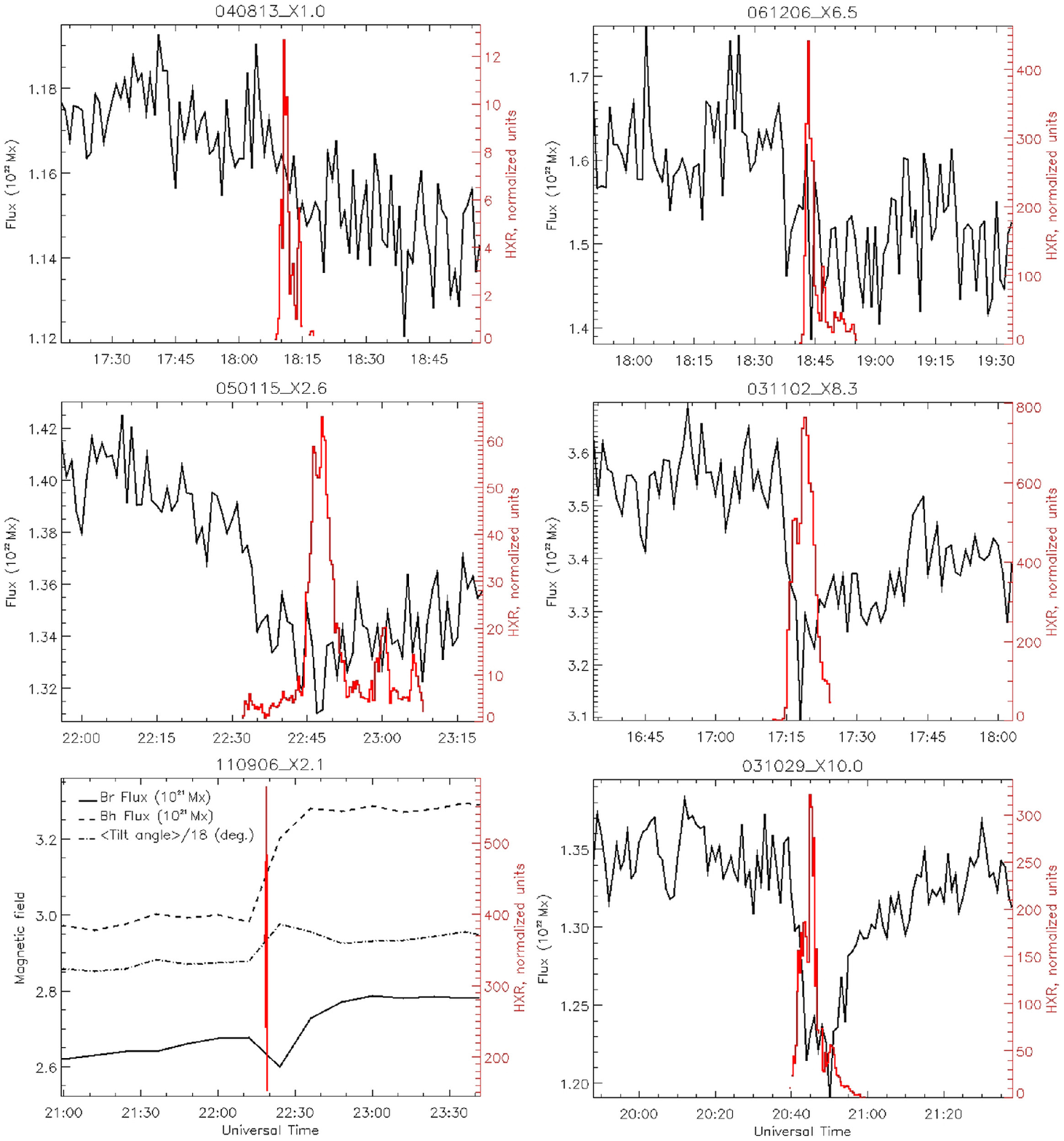}
\caption{Total unsigned flux as a function of time computed over a rectangular including the whole flared active region cropped to the area of the strongest field changes shown in Figures 2 and 3 for the five flares observed by GONG. The temporal evolution profile of the total flux in Br, Bh components of the vector field and average tilt angle over the region cropped to the area of the strongest field changes (2 degrees in latitude by 1 degree in longitude rectangular region around center of the region shown in Figure 4) for the flare observed by HMI is shown in the bottom left panel. The total HXR flux for each of the flares is over plotted in red.}
\label{default}
\end{center}
\end{figure}

As seen in Figures 2 and 3 the strongest field changes during flares occur in the regions with strong magnetic field close to the PIL, consistent with some of the previous studies \citep[e.g.,] [] {Sudol05,Zharkova05,Petrie10,Petrie12}. To look at the field changes at the PIL from a different perspective, we compute the difference between the average of the ten magnetograms immediately after GOES flare end time and the average of the ten magnetograms immediately before the GOES flare start time. GONG magnetograms of the five flaring active regions are shown in Figure 12, with the locations of the strongest magnetic field changes represented by 50 and 100 G contours of field change. Then we compute total unsigned flux in the area restricted by each of the contours. We select only regions showing a step-wise change in their total flux profile. These regions are labeled by a number.

\begin{figure}[h]
\begin{center}
\includegraphics[width=.95\textwidth]{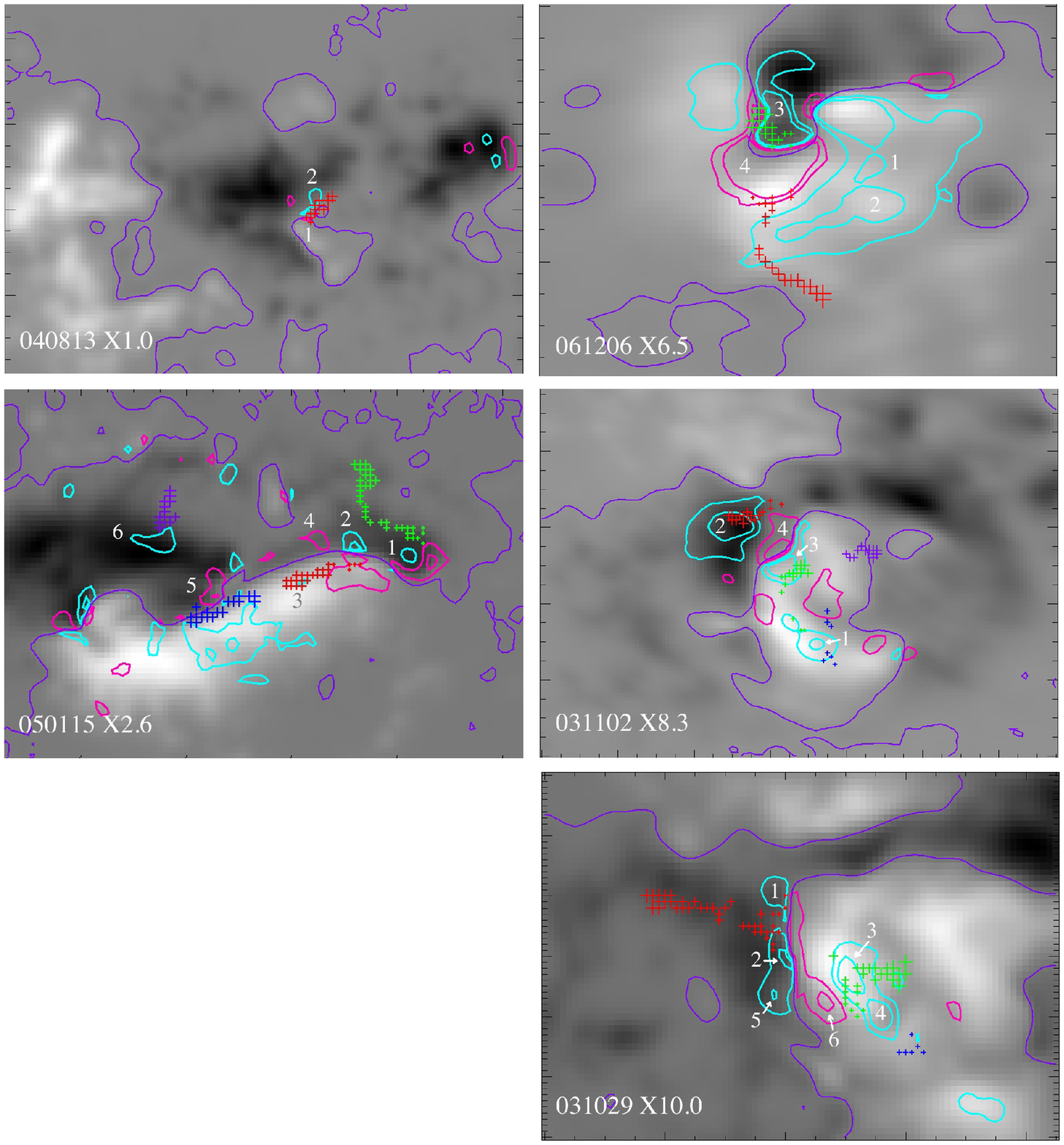}
\caption{GONG magnetograms of the five flaring active regions with the locations of the strongest magnetic field changes represented by 50 and 100 G contours of field change. Unsigned field decreases are light-blue, increases are magenta, and the PIL is the purple curves. The temporal evolution of the HXR centroids is shown with plus symbols, which increasing size represents progressing times during flares and different color denotes different sources. The contoured areas showing a stepwise change during flare are labeled with numbers.}
\label{default}
\end{center}
\end{figure}

We compute the midpoint time of the step for each of the numbered regions of field change shown in Figure 12. We also compute the midpoint time of the step in the total fluxes of the whole flared regions in Figure 11. The times are summarized in Table 2. 

\begin{table}[h]
\begin{center}
\caption{Midpoint times of the step-wise field changes and times of the HXR emission during the five flares observed by GONG. All times are in minutes from GOES flare start time.}
\begin{tabular}{lccccc}
\tableline\tableline
                           &       X1.0     &        X2.6     &       X6.5        &      X8.3      &      X10.0    \\  
\tableline
HXR interval    &   1.0-10.8   &   7.0-42.6   &   12.0-26.5   &   9.0-21.5   &   2.0-22.7   \\   
HXR peak        &         3.8       &      23.0       &        14.5       &        16.0     &        8.0       \\     
\tableline 
Total region     &         2.5       &        8.5       &           8.5       &        12.5     &        5.5       \\
\tableline
PIL region 1     &      11.5        &       6.5        &        12.5       &        11.5     &        2.5       \\
PIL region 2     &      12.5        &       9.5        &        13.5       &        13.5     &        4.5       \\
PIL region 3     &         -            &       9.5        &        14.5       &        13.5     &        4.5       \\
PIL region 4     &         -            &     20.5        &        20.5       &        17.5     &        5.5       \\
PIL region 5     &         -            &     24.5        &           -           &           -        &        8.5        \\
PIL region 6     &         -            &     30.5        &           -           &           -         &     13.5        \\ 
\tableline
\end{tabular}
\end{center}
\end{table}

Comparing locations of the numbered regions and HXR sources in Figure 12 with the midpoint times of the regions from Table 2, we see that in most of the flares the flux changes progress along the PIL from the initial position of the HXR sources in the direction of the HXR motion. They propagate first along and then away from the PIL. As we discussed in Section 5.2, we also find that some of the field changes follow the same trajectory with a similar speed as the HXR footpoint but at an earlier time (see Figure 9). Comparing the timing and location of the stepwise field change regions near the PIL from Table 2 and Figure 12 with the timing of the field changes from Figure 9, we find that they generally agree. From Figure 9, the midpoint times of the field changes at the very early locations of the HXR footpoints seem to be very close to the HXR times. As the mid-step time is about a minute or more later than the start time of the step, we can conclude that the field changes start from about a minute to a few minutes earlier than the detectable HXR signal.       

The results from Figure 12 and Table 11 also confirm that the midpoint times of the field increases appear to be later in comparison with the field decreases, in agreement with the results discussed in Section 5.2. For some of the regions, the later midpoint times of the field changes are more likely due to a longer step rather than a later start of the field changes. This does not look to be the case for the regions where the field decreases. The start of the step from the decreasing flux profiles, like the midpoint of the step, indeed seems to occur at a progressively later time in the direction of the HXR footpoint motion.

\section{Discussion}
We investigate the spatial and temporal connection between photospheric magnetic field changes during six X-class flares and the location of HXR emission observed by RHESSI to relate the field changes to the coronal restructuring and investigate the origin of the field changes. We interpret these field changes as a response to magnetic reconnection in the corona. Although we do not observe the reconnection process itself, the observed spatio-temporal relationship between the footprint HXR and the photospheric field changes indicates a relationship between the two, and the HXR is widely associated with reconnection. Therefore it is natural to associate the photospheric field changes to the reconnection also \citep[see, e.g.,][]{Temmer07,Jing08,Yang11}, even though the precise nature of this relationship is not yet clear. 

We find that spatially the early RHESSI emission corresponds well to locations of the strong photospheric field changes. The strong field changes occur predominantly in the regions of strong magnetic field near the PIL. The later RHESSI emission does not correspond to significant field changes as the flare footpoints are moving away from the PIL. Most of the field changes start before or around the start time of the detectable HXR signal, and they end at about the same time or later than the detectable HXR flare emission. Some of the field changes propagate with speed close to that of the HXR footpoint at a later phase of the flare, often after the HXR signal reaches its maximum, when the footpoint starts moving away from the PIL, i.e. the field changes follow the same trajectory as the HXR footpoint, but at an earlier time. Thus, the field changes and HXR emission are spatio-temporally related but not simultaneous. We also find that in the strongest X-class flares amplitudes of the field changes peak a few minutes earlier than the peak of the HXR signal.

Because of the several-minute time delays between the HXR peaks and the largest flux changes in some of the flares, and the sub-minute coronal travel times, a simple causal link between the HXR and flux changes does not seem to be possible. The HXR emission and the field-change effects are not likely to be an indication of the same event, though they must be related in some way. They may be a result of the consecutive process: flare onset, then photospheric magnetic field changes, then particles acceleration resulting in HXR emission. The main photospheric field changes may be due to earlier reconnection lower in the atmosphere, in contrast to the reconnection higher in the corona associated with the HXR peak. However if the main field changes were due to reconnection low in the atmosphere occurring earlier than the HXR peak, the HXR could not lag the main photospheric field changes along common spatial paths as observed. Instead, the HXR peak would be expected at a later time and in more distant location of the HXR source from the PIL, than the largest field changes. 

The Alfv\'{e}n speed in the corona is usually estimated to be about 1000 km/s, and can be as high as a few tenths of the speed of light in the lower corona in presence of strong, $\sim$500G and higher, magnetic fields \citep[see, e.g.,][]{Fletcher08}. An Alfv\'{e}n wave should take only a few minutes to cross an active region, and a fraction of a minute to get from a flare loop apex to the base of the corona transporting flare energy and causing the abrupt magnetic field changes. The accelerated particles would also have travel times much less than a minute. Thus these travel times cannot explain time delays as large as the several-minute gaps between the magnetic and HXR peaks for the largest flares in our analysis.

We suggest that the major magnetic field changes occur before most of the particle acceleration and HXR emission. For example, the formation of a current sheet at a magnetic X-point above a flaring loop system could produce major magnetic changes before the major particle acceleration phase. This explanation does not conflict with flux rope eruption models, including those involving ideal magnetohydrodynamical (MHD) instabilities and loss of flux-rope equilibrium. See, e.g., the two- and three-dimensional models of \cite{Lin00} and \cite{Fan07}, in which current sheets naturally form below the erupting flux ropes. Current sheets lead naturally to reconnection, though the physics is complex and unpredictable \citep[see, e.g.,][for simulations and observations]{Magara99,Drake06,Lin09}. \cite{Shibata11} make a statement that current sheet formation is a necessary condition for producing a flare. Thus we suggest, that the gap between the major flux change and the main HXR emission could correspond to the gap between the current sheet formation and the main reconnection, i.e., the HXR flare. 

Immediately after flare onset, the field changes would correspond to the current sheet formation phase, when the arcade field below the X-type neutral point collapses into a vertical sheet with Y-shaped structure at each end, and the low-lying flare loop legs form below the sheet. The timescales of the current sheet formation are likely close to Alfv\'{e}nic because the Alfv\'{e}n speed is the characteristic speed of the bulk dynamics in the corona. The high conductivity of the corona would preserve the original topology until the current sheet forms. At a certain stage of the current sheet formation, field derivatives would be large and resistivity would become important within the sheet, and reconnection would occur spontaneously. In this scenario, the field changes and the particle acceleration would originate from a common coronal site, but the few-minutes gap between the magnetic and HXR peaks would represent the observed pause between current sheet formation and reconnection. Also, the peak HXR must be emitted at the same loop's footpoints as the largest field changes, but later than these changes, and the peak HXR emission must correspond to ongoing reconnection of these flare loops \citep{Bogachev05}. This picture might also explain why HXR sources do not appear farther away from neutral line than the largest changes in the magnetic field.

The above discussion might help to explain why the most intense HXR emission tends to lag the largest field changes in the largest flares studied here, but this hypothesis needs to be tested. It is consistent with two reconnection sites moving vertically apart from each other as the flare develops. In radio observations, one can see a flare starting at some height in the corona, and then propagating both upward and downward \citep[e.g.,] [] {Altyntsev99}. Such radio sources correspond to the formation of a current sheet, with a reconnection site at the top and bottom. It would therefore be very interesting to determine whether the timing of radio source motions matches the picture described above, i.e., whether the separation of the radio sources coincides with the main field changes and comes before the peak HXR emission. The tearing mode instability and formation of secondary magnetic islands \citep[see, e.g.,][]{Drake06,Lin09} in the current sheet could be a reason for formation of multiple reconnection sites.  

Independent of the scenario described above, the fact that during some large flares the major flux change appears significantly earlier than the peak HXR emission, earlier than can be explained by simple causal arguments, provides a new observational constraint for flare models that may help us towards a better understanding of the complex and mysterious flare phenomenon.

\section*{Acknowledgments}

We thank Hugh Hudson for discussions and helpful comments. This work utilizes the Global Oscillation Network Group (GONG) data obtained by the NSO Integrated Synoptic Program (NISP), managed by the National Solar Observatory, which is operated by AURA, Inc. under a cooperative agreement with the National Science Foundation. The data were acquired by instruments operated by the Big Bear Solar Observatory, High Altitude Observatory, Learmonth Solar Observatory, Udaipur Solar Observatory, Instituto de Astrof'sica de Canarias, and Cerro Tololo Interamerican Observatory. The HMI and AIA data supplied courtesy of the SDO/HMI and SDO/AIA consortia. SDO is a mission for NASA's Living With a Star (LWS) Program. The Transition Region and Coronal Explorer (TRACE) is a mission of the Stanford-Lockheed Institute for Space Research, and part of the NASA Small Explorer program. O. Burtseva, G. J. D. Petrie and A. A. Pevtsov are partially supported by NASA grant NNX14AE05G. J. C. Martinez-Oliveros was supported by NASA grant NNX11AP05G for RHESSI.

\bibliography{Burtseva_etal_bib}

\begin{thebibliography}{}
\expandafter\ifx\csname natexlab\endcsname\relax\def\natexlab#1{#1}\fi

\bibitem[{{Altyntsev} {et~al.}(1999){Altyntsev}, {Grechnev}, {Nakajima},
  {Fujiki}, {Nishio}, \& {Prosovetsky}}]{Altyntsev99}
{Altyntsev}, A.~T., {Grechnev}, V.~V., {Nakajima}, H., {et~al.} 1999, \aaps,
  135, 415

\bibitem[{{Alvarado-G{\'o}mez} {et~al.}(2012){Alvarado-G{\'o}mez},
  {Buitrago-Casas}, {Mart{\'{\i}}nez-Oliveros}, {Lindsey}, {Hudson}, \&
  {Calvo-Mozo}}]{Alvarado12}
{Alvarado-G{\'o}mez}, J.~D., {Buitrago-Casas}, J.~C.,
  {Mart{\'{\i}}nez-Oliveros}, J.~C., {et~al.} 2012, \solphys, 280, 335

\bibitem[{{Asai} {et~al.}(2004){Asai}, {Yokoyama}, {Shimojo}, {Masuda},
  {Kurokawa}, \& {Shibata}}]{Asai04}
{Asai}, A., {Yokoyama}, T., {Shimojo}, M., {et~al.} 2004, \apj, 611, 557

\bibitem[{{Bogachev} {et~al.}(2005){Bogachev}, {Somov}, {Kosugi}, \&
  {Sakao}}]{Bogachev05}
{Bogachev}, S.~A., {Somov}, B.~V., {Kosugi}, T., \& {Sakao}, T. 2005, \apj,
  630, 561

\bibitem[{{Burtseva} \& {Petrie}(2013)}]{Burtseva13}
{Burtseva}, O., \& {Petrie}, G. 2013, \solphys, 283, 429

\bibitem[{{Cheng} {et~al.}(2012){Cheng}, {Kerr}, \& {Qiu}}]{Cheng12}
{Cheng}, J.~X., {Kerr}, G., \& {Qiu}, J. 2012, \apj, 744, 48

\bibitem[{{Cliver} {et~al.}(2012){Cliver}, {Petrie}, \& {Ling}}]{Cliver12}
{Cliver}, E.~W., {Petrie}, G.~J.~D., \& {Ling}, A.~G. 2012, \apj, 756, 144

\bibitem[{{Dennis} \& {Pernak}(2009)}]{Dennis09}
{Dennis}, B.~R., \& {Pernak}, R.~L. 2009, \apj, 698, 2131

\bibitem[{{Drake} {et~al.}(2006){Drake}, {Swisdak}, {Schoeffler}, {Rogers}, \&
  {Kobayashi}}]{Drake06}
{Drake}, J.~F., {Swisdak}, M., {Schoeffler}, K.~M., {Rogers}, B.~N., \&
  {Kobayashi}, S. 2006, \grl, 33, 13105

\bibitem[{{Fan} \& {Gibson}(2007)}]{Fan07}
{Fan}, Y., \& {Gibson}, S.~E. 2007, \apj, 668, 1232

\bibitem[{{Fisher} {et~al.}(2012){Fisher}, {Bercik}, {Welsch}, \&
  {Hudson}}]{Fisher12}
{Fisher}, G.~H., {Bercik}, D.~J., {Welsch}, B.~T., \& {Hudson}, H.~S. 2012,
  \solphys, 277, 59

\bibitem[{{Fletcher} \& {Hudson}(2002)}]{Fletcher02}
{Fletcher}, L., \& {Hudson}, H.~S. 2002, \solphys, 210, 307

\bibitem[{{Fletcher} \& {Hudson}(2008)}]{Fletcher08}
---. 2008, \apj, 675, 1645

\bibitem[{{Harker} \& {Pevtsov}(2013)}]{Harker13}
{Harker}, B.~J., \& {Pevtsov}, A.~A. 2013, \apj, 778, 175

\bibitem[{{Hoeksema} {et~al.}(2014){Hoeksema}, {Liu}, {Hayashi}, {Sun},
  {Schou}, {Couvidat}, {Norton}, {Bobra}, {Centeno}, {Leka}, {Barnes}, \&
  {Turmon}}]{Hoeksema14}
{Hoeksema}, J.~T., {Liu}, Y., {Hayashi}, K., {et~al.} 2014, \solphys, 289, 3483

\bibitem[{{Hudson} {et~al.}(2008){Hudson}, {Fisher}, \& {Welsch}}]{Hudson08}
{Hudson}, H.~S., {Fisher}, G.~H., \& {Welsch}, B.~T. 2008, in Astronomical
  Society of the Pacific Conference Series, Vol. 383, Subsurface and
  Atmospheric Influences on Solar Activity, ed. R.~{Howe}, R.~W. {Komm}, K.~S.
  {Balasubramaniam}, \& G.~J.~D. {Petrie}, 221

\bibitem[{{Ji} {et~al.}(2007){Ji}, {Huang}, \& {Wang}}]{Ji07}
{Ji}, H., {Huang}, G., \& {Wang}, H. 2007, \apj, 660, 893

\bibitem[{{Jing} {et~al.}(2008){Jing}, {Chae}, \& {Wang}}]{Jing08}
{Jing}, J., {Chae}, J., \& {Wang}, H. 2008, \apjl, 672, L73

\bibitem[{{Johnstone} {et~al.}(2012){Johnstone}, {Petrie}, \&
  {Sudol}}]{Johnstone12}
{Johnstone}, B.~M., {Petrie}, G.~J.~D., \& {Sudol}, J.~J. 2012, \apj, 760, 29

\bibitem[{{Kosovichev} \& {Zharkova}(2001)}]{Kosovichev01}
{Kosovichev}, A.~G., \& {Zharkova}, V.~V. 2001, \apjl, 550, L105

\bibitem[{{Krucker} {et~al.}(2003){Krucker}, {Hurford}, \& {Lin}}]{Krucker03}
{Krucker}, S., {Hurford}, G.~J., \& {Lin}, R.~P. 2003, \apjl, 595, L103

\bibitem[{{Lin} \& {Forbes}(2000)}]{Lin00}
{Lin}, J., \& {Forbes}, T.~G. 2000, \jgr, 105, 2375

\bibitem[{{Lin} {et~al.}(2009){Lin}, {Li}, {Ko}, \& {Raymond}}]{Lin09}
{Lin}, J., {Li}, J., {Ko}, Y.-K., \& {Raymond}, J.~C. 2009, \apj, 693, 1666

\bibitem[{{Liu} {et~al.}(2010){Liu}, {Lee}, {Jing}, {Liu}, {Deng}, \&
  {Wang}}]{Liu10}
{Liu}, C., {Lee}, J., {Jing}, J., {et~al.} 2010, \apjl, 721, L193

\bibitem[{{Magara} \& {Shibata}(1999)}]{Magara99}
{Magara}, T., \& {Shibata}, K. 1999, \apj, 514, 456

\bibitem[{{Masuda} {et~al.}(2001){Masuda}, {Kosugi}, \& {Hudson}}]{Masuda01}
{Masuda}, S., {Kosugi}, T., \& {Hudson}, H.~S. 2001, \solphys, 204, 55

\bibitem[{{Maurya} {et~al.}(2012){Maurya}, {Vemareddy}, \&
  {Ambastha}}]{Maurya12}
{Maurya}, R.~A., {Vemareddy}, P., \& {Ambastha}, A. 2012, \apj, 747, 134

\bibitem[{{Metcalf} {et~al.}(2003){Metcalf}, {Alexander}, {Hudson}, \&
  {Longcope}}]{Metcalf03}
{Metcalf}, T.~R., {Alexander}, D., {Hudson}, H.~S., \& {Longcope}, D.~W. 2003,
  \apj, 595, 483

\bibitem[{{Petrie}(2012)}]{Petrie12}
{Petrie}, G.~J.~D. 2012, \apj, 759, 50

\bibitem[{{Petrie}(2013)}]{Petrie13}
---. 2013, \solphys, 287, 415

\bibitem[{{Petrie}(2014)}]{Petrie14}
---. 2014, \solphys, 289, 3663

\bibitem[{{Petrie} \& {Sudol}(2010)}]{Petrie10}
{Petrie}, G.~J.~D., \& {Sudol}, J.~J. 2010, \apj, 724, 1218

\bibitem[{{Press} {et~al.}(2007){Press}, {Teukolsky}, {Vetterling}, \&
  {Flannery}}]{Press07}
{Press}, W.~H., {Teukolsky}, S.~A., {Vetterling}, W.~T., \& {Flannery}, B.~P.
  2007, {Section 15.5 Nonlinear Models, Numerical recipes: the art of
  scientific computing (3rd ed.)} (New York: Cambridge University Press.)

\bibitem[{{Qiu} \& {Gary}(2003)}]{Qiu03}
{Qiu}, J., \& {Gary}, D.~E. 2003, \apj, 599, 615

\bibitem[{{Shibata} \& {Magara}(2011)}]{Shibata11}
{Shibata}, K., \& {Magara}, T. 2011, Living Reviews in Solar Physics, 8, 6

\bibitem[{{Sudol} \& {Harvey}(2005)}]{Sudol05}
{Sudol}, J.~J., \& {Harvey}, J.~W. 2005, \apj, 635, 647

\bibitem[{{Svestka} {et~al.}(1982){Svestka}, {Hoyng}, {van Tend}, {Boelee}, {de
  Jager}, {Stewart}, {Acton}, {Bruner}, {Gabriel}, {Rapley}, {de Jager},
  {LaFleur}, {Nelson}, {Simnett}, {van Beek}, \& {Wagner}}]{Svestka82}
{Svestka}, Z., {Hoyng}, P., {van Tend}, W., {et~al.} 1982, \solphys, 75, 305

\bibitem[{{Temmer} {et~al.}(2007){Temmer}, {Veronig}, {Vr{\v s}nak}, \&
  {Miklenic}}]{Temmer07}
{Temmer}, M., {Veronig}, A.~M., {Vr{\v s}nak}, B., \& {Miklenic}, C. 2007,
  \apj, 654, 665

\bibitem[{{Tomczak} \& {Ciborski}(2007)}]{Tomczak07}
{Tomczak}, M., \& {Ciborski}, T. 2007, \aap, 461, 315

\bibitem[{{Wang}(1992)}]{Wang92}
{Wang}, H. 1992, \solphys, 140, 85

\bibitem[{{Wang} {et~al.}(1994){Wang}, {Ewell}, {Zirin}, \& {Ai}}]{Wang94}
{Wang}, H., {Ewell}, Jr., M.~W., {Zirin}, H., \& {Ai}, G. 1994, \apj, 424, 436

\bibitem[{{Wang} \& {Liu}(2010)}]{Wang10}
{Wang}, H., \& {Liu}, C. 2010, \apjl, 716, L195

\bibitem[{{Wang} {et~al.}(2002){Wang}, {Spirock}, {Qiu}, {Ji}, {Yurchyshyn},
  {Moon}, {Denker}, \& {Goode}}]{Wang02}
{Wang}, H., {Spirock}, T.~J., {Qiu}, J., {et~al.} 2002, \apj, 576, 497

\bibitem[{{Wang} {et~al.}(2012){Wang}, {Liu}, {Liu}, {Deng}, {Liu}, \&
  {Wang}}]{Wang12}
{Wang}, S., {Liu}, C., {Liu}, R., {et~al.} 2012, \apjl, 745, L17

\bibitem[{{Yang} {et~al.}(2011){Yang}, {Cheng}, {Krucker}, \& {Hsieh}}]{Yang11}
{Yang}, Y.-H., {Cheng}, C.~Z., {Krucker}, S., \& {Hsieh}, M.-S. 2011, \apj,
  732, 15

\bibitem[{{Yurchyshyn} {et~al.}(2004){Yurchyshyn}, {Wang}, {Abramenko},
  {Spirock}, \& {Krucker}}]{Yurchyshyn04}
{Yurchyshyn}, V., {Wang}, H., {Abramenko}, V., {Spirock}, T.~J., \& {Krucker},
  S. 2004, \apj, 605, 546

\bibitem[{{Zharkova} {et~al.}(2005){Zharkova}, {Zharkov}, {Ipson}, \&
  {Benkhalil}}]{Zharkova05}
{Zharkova}, V.~V., {Zharkov}, S.~I., {Ipson}, S.~S., \& {Benkhalil}, A.~K.
  2005, Journal of Geophysical Research (Space Physics), 110, 8104

\end{thebibliography}

\end{document}